\documentclass{ws-ijmpd}
\usepackage[super,compress]{cite}
\begin{document}

\def\RR{{\rm I\kern-.17em R}}
\def\eqq{\stackrel{S}{=}}

\markboth{Filipe C. Mena}
{CYLINDRICALLY SYMMETRIC MODELS OF GRAVITATIONAL COLLAPSE}

%
\catchline{}{}{}{}{}
%

\title{CYLINDRICALLY SYMMETRIC MODELS OF GRAVITATIONAL COLLAPSE TO BLACK HOLES: A SHORT REVIEW}

\author{FILIPE MENA}

\address{Centro de Matem\'atica, Universidade do Minho\\
4710-057 Braga, Portugal.\\
fmena@math.uminho.pt}



\maketitle

\begin{history}
\received{Day Month Year}
\revised{Day Month Year}
\end{history}

\begin{abstract}
We survey results about exact cylindrically symmetric models of gravitational collapse in General Relativity. We focus on models which result from the matching of two spacetimes having collapsing interiors which develop trapped surfaces and vaccum exteriors containing gravitational waves. We collect some theorems from the literature which help to decide {\em a priori} about eventual spacetime matchings. We revise, in more detail, some toy models which include some of the main mathematical and physical issues that arise in this context, and compute the gravitational energy flux through the matching boundary of a particular collapsing region. Along the way, we point out several interesting open problems. 
\end{abstract}


\ccode{PACS numbers: 04.20.Jb, 04.20.Dw, 04.30.Db, 04.40.Nr, 04.70.Bw} 

\section{Introduction}	
It is a consequence of Birkhoff's theorem that, in spherically symmetric vacuum spacetimes, there are no gravitational wave solutions to the Einstein Field Equations (EFEs). A next simplest symmetry assumption to model gravitational waves would be cylindrical symmetry. In fact, since it allows a $1+1$ decomposition of the EFEs, this type of symmetry seems to provide the only hope, in pratice, to construct exact models of interaction between matter and gravitational waves \cite{Bicak-review}.

In spite of being quite special, cylindrical symmetry, as an intermediate step towards the modelling of axially symmetric compact objects, has sucessfully been used to investigate a wide range of other physical situations, such as anisotropic stars \cite{Abbas}, rotating astrophysical objects \cite{Stephani}, cosmic strings  \cite{Vilenkin} and astrophysical jets \cite{Mashoon}.
%
It is also known that black strings form from a gravitational collapse of cylindrical dust clouds in the three-dimensional low-energy string theory \cite{Hyun}. 
This collapse process can also form naked singularities. 
In fact, cylindrical settings have been playing an important role in studies about the cosmic censorship conjecture \cite{Penrose}, the topological censorship conjecture \cite{Cai} and the hoop conjecture \cite{Hoop}. They have also been a natural framework to study fundamental issues in General Relativity (GR), such as the definition of gravitational energy \cite{Thorne}. 

This review is, by no means complete and, due to lack of space, deliberately excludes most of the models with collapsing (thin or thick) shells, alternative theories of gravity, higher dimensions and numerical approaches. Instead, we concentrate on collapsing exact solutions to the EFEs with fluid and scalar field sources, within (global) models where there is a matching to an appropriate exterior. Among those, we only give the details in a couple of examples which we use, as toy models, to illustrate the mathematical and physical problems that typically arise in this area of research.

The plan of the paper is as follows:  In Sections 2 and 3, we collect some general results about cylindrically symmetric collapse in GR. This material is scattered in the literature, partly contained in mathematical theorems, and we found useful to gather it here. We shall then revise some existing models of cylindrical gravitational collapse, in Section 4. The final Section 5, contains a short review, and some new computations, about the energy flux through the boundary of collapsing spatially homogeneous but anisotropic objects. 

Thoughout we use units such that $c=G=1$ and indices $\alpha, \beta=0,1,2,3$ and $i,j=1,2,3$.   
\section{Cylindrical symmetry and some consequences}
Cylindrically symmetric metrics in GR are surveyed in the excellent books \cite{Stephani, Griffith}. In this section, we recall some mathematical aspects, most of them not described in those books, which will be important for what follows. 
We start with a definition \cite{Carot}:
\begin{definition}
A spacetime is cylindrically symmetric if and only if it admits a Lie group $G_2$ on $S_2$ of isometries containing an axial symmetry.
\end{definition}
This definition implies that the $G_2$ group must be Abelian \cite{Barnes}. Under fairly general assumptions, an interesting geometric result about apparent horizon formation has been found \cite{Wang}:   
\begin{theorem}
Consider a cylindrically symmetric spacetime satisfying the dominant energy condition. If $\Lambda>0$, then there are neither outer nor degenerate apparent horizons.
If $\Lambda=0$, then there are no outer apparent horizons.
\end{theorem}
In this context, defining black holes by the existence of future outer apparent horizons or degenerate horizons, one concludes that there are no black holes in cylindrical symmetry for $\Lambda>0$. Black hole solutions with cylindrical symmetry have first been investigated in \cite{Lemos1} for $\Lambda<0$ (see \cite{MNT} for subsequent work). For $\Lambda=0$, an example of a "degenerate" cylindrical black hole can be found in \cite{Wang-scalar}. 

For the purpose of this paper, we will now assume that the $G_2$ group acts orthogonally transitively. In that case, the most general cylindrically symmetric line element reads \cite{Stephani}
\begin{equation}
\label{metric}
ds^2=e^{-2\psi}(e^{2\gamma}(-dT^2+d\rho^2) +R^2 d\varphi^2) +e^{2\psi}(dz+Ad\varphi)^2 
\end{equation}
where $\psi, \gamma$ and $R$ are $C^2$ functions of $t$ and $\rho$ to be determined for each matter content and the identification $\varphi +2\pi \to \varphi$ is assumed in order to get the correct topology. If the two Killing vectors are hypersurface orthogonal one can set $A=0$. 

Some solutions to the EFEs with the above line element do not contain an axis, and others, despite having an axis, have a metric which is not regular on it \cite{Stephani}. The condition which ensures the existence of a regular axis at $\rho=0$ is
\begin{equation}
\lim_{\rho\to 0} \frac{\nabla_\alpha (\xi_\beta \xi^\beta)\nabla^\alpha (\xi_\beta \xi^\beta)}{4\xi_\beta \xi^\beta}=1,
\end{equation}
where $\vec \xi$ is the axial Killing vector.   
Solutions with metric \eqref{metric} which do not necessarily contain an axis can still be interesting as they can act, for example, as outer regions for inner cylindrical sources. 

The notion of {\em asymptotic flatness} in cylindrical symmetry was studied in \cite{BCM}. Interestingly, it was found that the $2$-surfaces of transitivity are never trapped in electro-vacuum asymptotically flat spacetimes. This result applies to dynamical vacuum solutions as well and, as we shall see, has dramatic consequences on the existence of gravitational waves emitted from cylindrically shaped objects which collapse within their apparent horizon. 
\section{Spacetime matching and some consequences}
Let $(M^{\pm},g^{\pm})$ be spacetimes with non-null boundaries $S^{\pm}$.
Matching them requires an identification of the boundaries, i.e. a pair of embeddings
$\Phi^\pm:\; S \longrightarrow M^\pm$ with $\Phi^\pm(S) = S^{\pm}$,
where $S$ is an abstract copy of any of the boundaries. 
Let $\xi^i$ be a coordinate system on $S$. Tangent vectors to
$S^{\pm}$ are obtained by $f^{\pm \alpha}_i =
\frac{\partial \Phi_\pm^\alpha}{\partial \xi^i}$ though we shall usually work with orthonormal combinations $e^{\pm \alpha}_i$ of the $f^{\pm \alpha}_i$. There are also  unique (up to 
orientation) unit normal vectors $n_{\pm}^{\alpha}$ to the boundaries. 
We choose them so that if $n_{+}^{\alpha}$ points into $M^{+}$ then
$n_{-}^{\alpha}$ points out of $M^{-}$ or viceversa. 
The first and second fundamental forms
are simply \cite{Mars-Seno}
$$q_{ij}^{\pm}= e^{\pm \alpha}_i e^{\pm \beta}_j
g_{\alpha\beta}|_{_{S^\pm}},~~
H_{ij}^{\pm}=-n^{\pm}_{\alpha} e^{\pm
\beta}_i\nabla^\pm_\beta e^{\pm \alpha}_j.$$ 
\begin{theorem} The matching conditions (in the absence of shells), between two spacetimes $(M^\pm,g^\pm)$ across a non-null hypersurface $S$, 
are the equality of the first and second fundamental forms on $S^{\pm}$, i.e.
\begin{equation}
q_{ij}^{+}=q_{ij}^{-},~~~
 H_{ij}^{+}=H_{ij}^{-}.
\label{eq:backmc}
\end{equation}
%
\end{theorem}
When symmetries are present, one can choose the $e^{\pm \alpha}_i$  to reflect the symmetry so that $H^\pm_{ij}$ simplifies. This choice can be formalized in the following definition  \cite{Vera}:
\begin{definition}
If $(M^+,g^+)$ and $(M^-,g^-)$ 
both admit a $m$-dimensional group of symmetries, the final matched spacetime $(M,g)$ is said to preserve the symmetry $G_m$ if there exist
$m$-vectors on $S$ that are mapped by the push-forwards $d\Phi^\pm$ to the restrictions of the generators of $G_m$ to $S^\pm$, respectively.
\end{definition}
If there is an intrinsically 
distinguishable generator of $G_m$, such as an axial Killing vector, then the {\em matching preserving the symmetry} must ensure its identification at $S$. Here, we shall only consider matchings preserving the cylindrical symmetry, which is represented by an Abelian group $G_2$.

When the matching conditions are satisfied, the next result necessarily follows from the Israel conditions \cite{Mars-Seno}:
\begin{corollary}
\label{cor1}
If $(M^-,g^-)$ contains a perfect fluid with pressure $p$ and $(M^+,g^+)$ contains vacuum, then $p |_{_{S^-}}= 0$. In particular,
if $(M^-,g^-)$ is spatially homogeneous then $p=0$ everywhere.
\end{corollary}
Regarding the formation of trapped cylinders, the matching conditions provide a quite important result (an analogue result in spherical symmetry can be found in \cite{Fayos}):
\begin{corollary}
\label{cor2}
If $(M^-,g^-)$ develops non-singular trapped cylinders up to $S$, then $(M^+,g^+)$ must also contain trapped cylinders which match the interior ones on $S$.
\end{corollary}
This result is related to the continuity of the mass through the boundary and is quite useful to exclude some possibilities of spacetime matching, for example with asymptotically flat non-stationary electro-vacuum spacetimes, since these do not contain trapped cylinders \cite{BCM}. We now combine the previous corollary with the results of \cite{BCM} to write:
\begin{theorem} 
\label{theo-waves}
Let $(M^+,g^+)$ be cylindrically symmetric and asymptotically flat, according to \cite{BCM}. If a collapsing cylindrical spacetime $(M^-,g^-)$ develops trapped cylinders, then $(M^+,g^+)$ cannot be non-stationary electro-vacuum. In particular, $(M^+,g^+)$ cannot be a vaccum exterior with gravitational waves.    
\end{theorem}
It is then clear that, if an interior spacetime contains trapped surfaces, one should look, whenever the case, for gravitational wave vacuum exteriors which are not asymptotically flat. 

We end this section, recalling a practical aspect for the matching procedure that we will use: Suppose that the interior contains a matter fluid. Then, the timelike boundary of the interior must be ruled by matter world-lines, which for dust are geodesics, and the matching conditions require that the boundary is also ruled by timelike geodesics of the vacuum exterior. If the interior is not dust, then the matter world-lines ruling the boundary are accelerating and now the accelerations at both sides of the boundary must match. 

These considerations allows one to make some clever choices regarding the parametrization of the matching (boundary) surface, and thus simplify the calculations. 
They also make more transparent the setting from the point of view of differential equations. In fact, one can formulate the problem as a {\em Cauchy problem} in the sense that, given an interior spacetime and "initial" conditions at the timelike boundary (provided by the matching), one may ask whether an exterior solution exists. Furthermore, one can split the question into {\em local} existence and {\em global} existence of solutions. One may also ask if the exterior solution is unique which, in turn, may have interesting physical consequences. Examples of results along those lines
 can by found in \cite{Mars-Seno-unique, Vera-unique}.

\section{Models of gravitational collapse}
In this section, we revise several cylindrically symmetric models of gravitational collapse. We shall put emphasis on what we can or cannot model which can be translated into which spacetimes can or cannot be matched. 

Motivated by the modelling of isolated regions, we mainly focus on vacuum outer regions. Nevertheless, outer regions with other sources have also been considered in the past and we survey some of them in sub-section \ref{other-exteriors}. We shall also concentrate on interiors with perfect fluid matter, as these have already a rich structure which allow us to point out the main issues. Other interiors are reviewed in sub-section \ref{other-interiors}.

\subsection{Vacuum exteriors}
 
The cylindrically symmetric static $\Lambda$-vacuum spacetime was discovered by Linet and Tian \cite{L,T}. That spacetime includes the $\Lambda=0$ particular case which is the well known Levi-Civita spacetime \cite{Levi}.  In turn, the cylindrical stationary $\Lambda$-vacuum metric was found by van Stockum \cite{Stock}, while the $\Lambda=0$ sub-case is the Lewis spacetime \cite{Lewis}. 
None of the above static or stationary $\Lambda$-vacuum cylindrical spacetimes contain trapped cylinders so, by Corollary \ref{cor2}, they are excluded as potential outer regions for collapsing interiors which develop trapped cylinders.

We shall then consider non-stationary vacuum exteriors $(M^+,g^+)$. In that case, the most general metric form is still
written as \eqref{metric}
and, now, interpreted as modelling the propagation of cylindrical gravitational
waves, see \cite{BCM}. For $A\ne 0$, the two Killing vectors are
not hypersurface orthogonal and the gravitational
waves have two polarisations states. One of the EFEs is
\[ R_{TT}-R_{\rho\rho}=0, \] which has the general
solution 
\begin{equation}
 \label{Rext} R(T,\rho)=F(x)+G(y), 
 \end{equation} where
$x=T+\rho,\;\;y=T-\rho$ and the subscripts denote partial differentiation.
Metric (\ref{metric}) is invariant under a
coordinate transformation 
\begin{equation}
 \tau:(x,y)\to(f(x),g(y)),
\label{tau} 
\end{equation}
 where $f$ and $g$ are arbitrary differentiable
functions (with non-zero derivative). 

As far as black hole formation is concerned, we will need to check whether the $2$-surfaces of
transitivity are trapped or marginally-trapped. From (\ref{metric}),
we see that such $2$-surfaces 
are trapped if \cite{Tod-Mena}
\begin{equation}
4R_xR_y=(R_T-R_{\rho})(R_T+R_{\rho})\geq 0 \label{TS}, 
\end{equation}
and
marginally-trapped if this expression is zero. It follows
from (\ref{Rext}) that if $R_x=0$, at say $x=x_0$, then all
$2$-surfaces of transitivity with $x=x_0$ are marginally-trapped.
Similar statements follow with $y$ replacing $x$.

Due to Theorem \ref{theo-waves}, the requirement of asymptotic flatness
implies that the $2$-surfaces of transitivity are never trapped. Specifically, in this case, \cite{BCM} showed that using the freedom \eqref{tau} one can choose a comoving radius of
the form $R(T,\rho)=\rho$, which forbids trapping. The choices $A=0$ and $R=\rho$ in (\ref{metric}) lead to the well-known Einstein-Rosen metric \cite{Einstein-Rosen} which, therefore, won't be suitable for our purposes. 

Instead of using \eqref{tau} to choose $R$, we will use \eqref{tau} to prescribe the matching surface $S^+$ and, therefore, finding $R$ becomes part of our problem. 
\subsection{Other exteriors}
\label{other-exteriors}

Null dust solutions on anti-de-Sitter backgrounds, as given by {\em Robinson-Trautman metrics} (sometimes also refered as {\em modified Vaidya metrics}),
 were considered in \cite{Senovilla} as:
\begin{equation}
\label{vaidya}
ds^2 = -\left(\alpha^2 \rho^2-\frac{4m(v)}{\alpha \rho}\right)dv^2+2dv d\rho +\rho^2(d\theta^2+dz^2)
\end{equation}
for
$$
T_{\alpha\beta}=\frac{4 m'}{\alpha \rho^2} k_\alpha k_\beta
$$
where $k_\alpha=-\delta^{~v}_\alpha,  k_\alpha k^\alpha=0$, $\alpha=\sqrt{-\Lambda/3}$, $m(v)$ represents the mass per unit lenght and $m'$ its derivative. 
 These solutions describe a flow of unpolarised radiation in the geometrical optics approximation. Besides modelling radiation in the form of neutrinos or electromagnetic waves, these solutions can also represent gravitational radiation. In \cite{Senovilla}, they are shown to describe (i) the formation
 of cylindrical black holes through the collapse of pure monochromatic light; (ii) the transformation of
a naked singularity into a black hole by reception of light; (iii) the annihalation of a naked singularity by 
sending a fine-tuned amount of light. None of these scenarios requires spacetime matching. 
From the Israel conditions, it is clear that a dust spacetime cannot be an interior
 to \eqref{vaidya} with $m^{\prime}\ne 0$. So, it is worth trying to match these exteriors to more general interiors. An example, in this direction, was given in \cite{Sharif-Abbas} for a heat conducting charged anisotropic collapsing fluid. 

Electro-vacuum {\em static} exteriors were explored in \cite{Sharif-Abbas-charged} using the plane symmetric static metrics for Einstein-Maxwell fields \cite{Stephani}
\begin{equation}
ds^2 = -\left( \frac{4m}{\rho}-\frac{2q^2}{\rho^2}\right)dT^2+\left( \frac{4m}{\rho}-\frac{2q^2}{\rho^2}\right)^{-1}d\rho^2+\rho^2(d\theta^2+dz^2)
\end{equation}
where $m$ and $q$ represent the (constant) mass and charge per unit length, respectively. The interior was taken to be a charged collapsing perfect fluid. One of the conclusions, in this case, was that the Coulomb and gravitational forces balance each other on the matching boundary. Trapped surface formation is not studied in \cite{Sharif-Abbas-charged} though, so the study is incomplete. 

Soliton solutions, as particular cases of \eqref{metric}, have been revised in \cite{Verdaguer}. In that book, there is a full section dedicated to cylindrically symmetric solitons, in particular to Einstein-Rosen soliton metrics, two polarisation waves and Faraday rotation which would certainly be interesting to explore as exteriors to gravitational collapse.

\subsection{Perfect Fluid Interiors}

Here, we use perfect fluids as our exact models of collapse and, in most cases, we will stick to dust fluids as toy models since they provide explicit solutions.

\subsubsection{Spatially homogeneous and isotropic}  

Spatially homogeneous and isotropic collapsing spacetimes in cylindrical symmetry have been studied in \cite{Seno-Vera, Nolan-Nolan, MNT, MacCallum}. These are simply given by the classical Friedman-Lema\^itre-Robertson-Walker (FLRW) metrics, but this doesn't seem to be very well known. 

A remarkable result of \cite{Seno-Vera} forbids the cylindrical counterpart of the (spherically symmetric)
Oppenheimer-Snyder model of gravitational collapse, in the sense that it was found to be impossible to match non-static FLRW metrics to any static vacuum in cylindrical symmetry.   Moreover, a quite strong rigidity result by \cite{Mars} says that, if the FLRW matter-density is non-negative, then any {\em static vacuum} exterior to FLRW must be spherically symmetric. This result has been generalised to axially symmetric {\em stationary vacuum} spacetimes \cite{Nolan-Vera}, which are then also forbidden as exteriors to FLRW. More details about rigidity and uniqueness results, in this context, can be found in the review article \cite{MMV}.

If we insist on vacuum exteriors, this leaves us with the non-stationary cases as potential exteriors to collapsing FLRW. As we shall see in the next section, this kind of matching is possible but at the cost of having trapped surfaces in the exterior, extending back to the initial hypersurface, as in Figure \ref{fig}. 

\subsubsection{Spatially homogeneous and anisotropic} 

\label{homogeneous}
Spatially homogeneous but anisotropic collapsing spacetimes in cylindrical symmetry were investigated in \cite{MTV,MNT}. Due to the presence of an axis \cite{MTV}, the spacetimes are locally rotationally symmetric (LRS). 
Now, all LRS spatially homogeneous metrics can be written in the compact
form \cite{Stephani}
\begin{equation}
\label{compact}
ds^{2}=-dt^2+a(t)^2 \bm{\theta}^2+b(t)^2\left[(dr-\epsilon rdz)^2+F'(r)^2 d\varphi^2\right]
\end{equation}
where
\[
\label{theta1}
\bm{\theta}=dz+n(F(r)+k)d\varphi~~~~~{\text {and}}~~~~~
F(r)=
\left\{
\begin{array}{cl}
-\cos r,& k=+1\\
r^2/2, & k=0\\
\cosh r, & k=-1,
\end{array}
\right.
\]
the prime denotes differentiation, and $\epsilon$ and $n$ are given by 
\begin{equation}
\label{epsi}
\epsilon=0,1;~~n=0,1;~~\epsilon n=\epsilon k=0,
\end{equation}
which have different combinations of values for Kantowski-Sachs and the different Bianchi types, see a classification in \cite{MTV}. We note that \eqref{compact} contains the FLRW metrics as particular cases.

It is a consequence of the results of \cite{MTV} that non-vacuum spacetimes \eqref{compact} cannot have a static vacuum exterior. A generalisation of this result to stationary vacuum exteriors is yet to be proved. However, given Corolary \ref{theo-waves}, we do not expect, in general, to be able to have such exteriors, at least in the cases where interior trapped surfaces do form. So, one should look for non-stationary exteriors. In particular, we will now summarise the matching between the above LRS  spacetimes and the family of non-stationary vacuum metrics \eqref{metric}.

The construction begins with a choice of an interior solution which, given Corollary \ref{cor1}, must be dust. 
%
As a consequence, the
matching is  performed across a timelike hypersurface $S^-=\{t(\lambda), r(\lambda), \varphi, z\} $ ruled by
matter trajectories, which are geodesics with a parameter $\lambda$. 
Therefore, we are free to set $\dot t =1$ and $\dot r =0$. From
the point of view of the exterior, the matching surface can be written in terms of the
coordinates $x,y$ of (\ref{tau}) as $S^+=\{x(\lambda),y(\lambda), \varphi, z\}$. The product $\dot{x}\dot{y}$
does not vanish since the matching surface is everywhere timelike
and so we may use the transformation $\tau$ of (\ref{tau}) to
set $x=\lambda, y=\lambda$, or equivalently $\dot T=1$ and
$\dot\rho=0$. Now the matching surface is at $r=r_0, \rho=\rho_0$
and the time coordinates can be taken to agree, i.e. $t=T$.

The full matching conditions were written in \cite{Tod-Mena}. One of the consequence of those conditions is that $\epsilon=0$, excluding e.g. Bianchi type V which is a generalisation of the $k=-1$ FLRW metrics. We note that we are considering non-tilted fluids here
 and an example with a tilted Bianchi type V spacetime was given in \cite{MNT}.

Given an interior solution, the matching conditions provide data at the boundary for the exterior, i.e. the functions $R,\psi, \gamma$ and their normal derivatives.
The EFEs in the exterior are a system of
hyperbolic equations in the $2$-dimensional quotient space
${\cal{Q}}$ of the spacetime by the symmetries (this is the
$(T,\rho)$-space) and the data is given on a non-characteristic
curve in ${\cal{Q}}$. The boundary of the matter $\rho=\rho_0$ defines such a
curve, even though it is a timelike surface in the full
$4$-metric. Therefore, we can deduce existence and uniqueness of
solutions in the domain of dependence in ${\cal{Q}}$ of the curve
on which the data is given.

\begin{figure}[!htb]
\centerline{\def\epsfsize#1#2{0.7#1}\epsffile{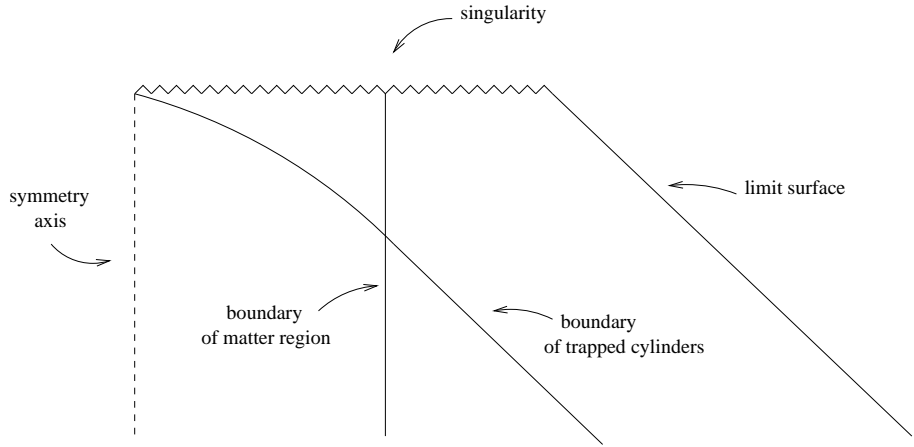}}
\caption{\label{fig} Schematic diagram of the spacetime
structure, as in \cite{Tod-Mena}.}
\end{figure}

 If the interior collapses to a
singularity, as it does for dust, then the data-curve has
an end at the singularity and there is a (one-dimensional) Cauchy horizon in the exterior in ${\cal{Q}}$,
which determines a limit surface ${\cal{H}}$ in the spacetime beyond which the solution is not determined by the data, see Figure \ref{fig}. Unfortunately, one expects\cite{Tod-Mena}
the exterior to be singular on this surface ${\cal{H}}$.

Given explicit interiors solutions to the EFEs, we may obtain $R$ explicity from the matching conditions. This is the case of Bianchi type I. In that case, it is straightforward to see that $R_y$ is always negative on
$S$ while $R_x$ necessarily changes sign. Thus, by
(\ref{TS}), there is always a marginally-trapped cylinder on
$S$ and this gives rise to a null hypersurface of
marginally-trapped cylinders ${\cal{N}}_0$. Cylinders to
the past of ${\cal{N}}_0$ are not trapped and those to the future
are. The main result of \cite{Tod-Mena} is summarised as:
\begin{theorem}
\label{theo-bianchi}
A cylindrical interior dust metric \eqref{compact}
which is diagonal (with $n=\epsilon=0$) can be
matched to a diagonal (with $A=0$) vacuum exterior \eqref{metric}. The exterior solution exists 
within the domain of dependence of the matching surface. The metric is
smooth up to the limit surface ${\cal{H}}$, and there are trapped
surfaces in the exterior that extend to the initial hypersurface. 
\end{theorem}
%
%
%
Although the condition $\epsilon=0$ is necessary for the matching, the condition $n=0$ is not. 
In principle, one could repeat the previous steps for 
 non-diagonal interior metrics (of Bianchi types II, VIII and IX)
matched to non-diagonal exterior vacuum metrics. This is hampered
by two things: the non-existence, in the literature at least, of
even implicit solutions for the interior metric; and the fact that
the EFEs are now coupled and cannot be solved one by one. This is then an open problem. Partial results for Bianchi type II can be found in \cite{Tod-Mena}.

\subsubsection{Inhomogeneous}
\label{inhomogeneous} 

The matching of collapsing perfect fluids to Einstein-Rosen waves were investigated in \cite{Herrera-Santos-Mac}. The authors noted that, since the interior radial pressure vanishes at the matching boundary, 
any local pressure effect, predicted by a local characterisation of the flux of the waves' gravitational 
energy, by means of pseudo-tensors, will not be observed. In the next section, we will come back to this issue and do an explicit calculation of the energy flux, in this context, using the solutions of the previous section as a toy model. 

The authors of  \cite{Morisawa, Sharif-Ahmad} also consider perfect fluid collapse, for the case of thick shells, and treat separately the case of dust \cite{Morisawa-dust}. In particular, they use a {\em high speed approximation} to work out interesting physics such as the effect of a bounce in the production of gravitational waves (which they also take as being Einstein-Rosen waves). In case the equation of state is sufficiently soft, they find that the collapse is maintained until a spacetime singularity forms. Interestingly, they conclude that higher collapsing speed produces more gravitational waves.   
 
None of the previous results uses interior {\em exact} solutions. In fact, there are very few known inhomogeneous cylindrically symmetric (non-stationary) exact solutions with perfect fluids, see \cite{Stephani} and \cite{Seno-Vera-1998, Seno-Vera-2001} and references therein.
For instance, the only known explicit dust metric containing an axis is due to \cite{Vera-Seno-2} and reads
\begin{equation}
\label{Senovilla-metric}
ds^2=-dt^2+dr^2+\left(1-\frac{t^2+r^2}{\alpha^2}\right)^2 dz^2+r^2d\varphi^2 .
\end{equation}
These spacetimes belong to the Szekeres class II family and are of Petrov type D. They contain a curvature singularity $t^2+r^2=\alpha^2$ which is spacelike in the region $r\in [0, \alpha/\sqrt 2[$, null for $r=\alpha/\sqrt 2$ and timelike for $r\in~ ]\alpha/\sqrt 2, \alpha]$. We are interested in the collapse to a spacelike singularity so we would like to think of \eqref{Senovilla-metric} with $r\in [0, \alpha/\sqrt 2[$ and look for a suitable exterior in the family of metrics \eqref{metric}. It is easy to check that there are trapped surfaces for $t^2\ge \alpha^2/2$, i.e. $r^2\le \alpha^2/2$, and these go all the way to $\{t=\alpha, r=0\}$. One can then cut the 
spacetime \eqref{Senovilla-metric} at a fixed radius and try to match it to a vacuum metric \eqref{metric} in the exterior. This can actually be done by methods similar to those described above for the spatially homogeneous metrics and the details will appear soon \cite{DMS}.  
\subsection{Other interiors}
\label{other-interiors}

Null dust collapse was studied in the articles \cite{Letelier-Wang, Nolan, Ghosh}, whose main purpose was to study the cosmic censorship conjecture. They find not only black hole formation, but also naked singularities as the final outcome of collapse, in those settings. Another study using null dust was done in \cite{Lemos-null} for collapsing shells of radiation in anti-de-Sitter backgrounds with the interesting conclusion that there is no naked singularity formation.

Imperfect (dissipative) fluids were investigated in \cite{Herrera} with further developments to include charge in \cite{Sarbari}. The effects of anisotropy and charge were analysed in those systems, however there are no results about trapped surface formation or matching to an exterior solution. 
This seems hard to study analytically, in part because there are no explicit solutions to the EFEs available, yet, for those cases. 

A paper generalising the perfect fluid collapse studied in \cite{Herrera-Santos, Herrera-Santos-Mac} can be found in \cite{Chakra}. In that paper, an anisotropic dissipative fluid is considered as the collapsing interior to the Einstein-Rosen exterior. The authors' main conclusion is that the radial pressure of the fluid is balanced by the shear viscosity, at the matching boundary $S$, resulting in a zero effective fluid pressure on $S$. In turn, the fact that the radial pressure is non-zero on $S$ might enable the existence of gravitational waves in the exterior. Although this study is interesting, it does not cover an analysis about the eventual gravitational energy flux through the matching boundary, or about trapped surface formation in the interior, which might lead to a breakdown in the spacetime matching as in \cite{Nolan-Nolan}. 

Charged Perfect fluids were studied in \cite{Sharif-Abbas-charged}. Interestingly, a new explicit exact solution is derived and used to model gravitational collapse which, in some cases, ends up in conical singularites. To complete the model, the authors match the interior fluid to an exterior charged {\em static}� cylindrically symmetric electro-vacuum solution.They find that the gravitational and Coulomb forces of the system balance each other on the matching surface. As the authors mention, it would be interesting to generalise the model to include electromagnetic radiation in the exterior. A study in this direction is in \cite{Sharif-Fatima}.

Scalar field spacetimes with a homothetic self-similarity were studied in \cite{Wang-scalar}, where a new class of
exact solutions to the Einstein-massless scalar field system was found. The paper includes a careful analysis of trapped surfaces and studies the cases where there is black hole formation in cylindrical symmetry. The global geometric properties of the spacetime are analysed but there is no matching to an exterior.  More recently, self-similar scalar fields with a non-minimal coupling and a regular axis were investigated in \cite{Condron1, Condron2} where, after a careful analysis of the spacetime global structure, a cosmic censorship result was established.
\section{Energy and gravitational waves}

Due to the well known difficulty in defining energy for the gravitational field in General Relativity, the definition of gravitational radiative flux measures in cylindrical symmetry is also problematic. Measures of gravitational radiation in cylindrical symmetry were revised in detail in \cite{MacCallum} together with their potential problems. 
Some measures that have been used in the past include the C-energy \cite{Thorne}, pseudo-tensors \cite{Landau}, Weyl spinors \cite{Hofmann} and the super-energy tensor \cite{Alfonso}. It is also worth to read \cite{Hayward} for a discussion about the conserved energy momentum for gravitational waves in cylindrical symmetry.

The authors of \cite{Morisawa} have used the C-energy in order to study the radiation flux during the collapse of inhomogeneous perfect fluid cylindrical shells of matter, using the metric (\ref{metric}). 

We shall give an example about how can this be done, using spatially homogeneous anisotropic spacetimes, for simplicity. 
We shall use the $C$-energy approach which, besides leading to a small but new result, allows a direct comparison with the results of \cite{Morisawa}. 


The energy flux associated to the C-energy potential, for the diagonal case of metric \eqref{metric}, reads \cite{Thorne}
\begin{equation}
\label{cflux}
C(T,\rho)=\frac{1}{8}(1-(R_\rho^2-R_T^2)e^{-2\gamma}).
\end{equation}
We recall that the Hawking mass for a closed surface $\Omega$ with surface area element $d\Omega$ is \cite{Herrera-Santos-Mac}
\begin{equation}
m=\frac{1}{(4\pi)^{3/2}}\left(\int_\Omega d\Omega \right)^{1/2}\left(2\pi-\int_\Omega (\nabla^\alpha k_\alpha^- \nabla^\beta k_\beta^+) d\Omega \right)
\end{equation}
where $k_\alpha^\pm$ are the outgoing and the ingoing null vectors orthogonal to $\Omega$. A simple calculation yields \cite{Thorne}
$$
\nabla^\alpha k_\alpha^- \nabla^\beta k_\beta^+=\frac{1}{8R^2}e^{2(\psi-\gamma)}(R_T^2-R_\rho^2),
$$
and, therefore, the quantity $C$ can also be regarded as a measure of the loss of Hawking mass per unit lenght in $z$. 
The energy flux associated to the C-energy (i.e. the energy per unit lenght in each cylinder of radius $\rho$) is
\begin{equation}
\label{jflux}
J^\alpha=\frac{e^{2(\psi-\gamma)}}{R}(C_{\rho}\delta_{T}^{~\alpha}-C_{T}\delta_\rho^{~\alpha}).
\end{equation}
We now consider the matching, described in Section \ref{homogeneous}, between the spatially homogeneous anisotropic perfect fluid diagonal metrics from \eqref{compact} and the diagonal vacuum metrics from \eqref{metric}. We can then substitute \eqref{cflux} in \eqref{jflux} and use the matching conditions from \cite{Tod-Mena} to derive very simple expressions, at the boundary $S$, for the energy flux as
\begin{eqnarray}
\label{jjj}
J^T &\eqq& \frac{F''}{a^3}(\dot a^2 -a^2 \mu) \\
J^\rho&\eqq&\nonumber \frac{\dot a^3 b+a\dot a^2\dot b}{a^4}
\end{eqnarray}
where $\mu$ stands for the matter-density.
So, the $C$-energy flux increases with the velocities of collapse $\dot a$ and $\dot b$. This is in agreement with the results of \cite{Morisawa} for perfect fluid shells of matter. In the case of dust 
$$
\mu(t)=2\frac{\dot a\dot b}{ab}+\frac{\dot b^2}{b^2}+\frac{k}{b^2}
$$
which can be readily substituted in \eqref{jjj}. In some cases, like the dust Bianchi I, where $F''=1$ and $k=0$, it is possible to obtain explicit solutions to the EFEs as
$$
a(t)=(\alpha-t)^{-1/3}(\beta-t)~~~~~\text{and}~~~~~b(t)=(\alpha-t)^{2/3},
$$
where $t<\text{min} \{\alpha, \beta\}$, with $\alpha, \beta \in \RR^+$, and compute \eqref{jjj} explicitely.
\section{Concluding remarks}

Numerical GR and advances in computer technology have contributed to huge progress on models of gravitational waves generation from isolated objects in astrophysics. These approaches rely mainly 
on linearising the EFEs and looking at gravitational radiation as a {\em linear} perturbative effect on otherwise highly symmetric backgrounds. However, since GR is fundamentaly a {\em nonlinear} theory, exact nonlinear models of gravitational waves can give important physical and mathematical insights to this problem. Although simplistic, and often unrealistic (e.g. if they impose infinitely long cylinders), cylindrically symmetric models of gravitational collapse have been used as an important tool, since they can capture interesting nonlinear physical effects \cite{Bicak-review}. 

Mathematically, cylindrical symmetry implies some no-go results, as revised above, which are partly due to the peculiar topology that is being imposed. Nevertheless, it is the simplest natural symmetric setting to model gravitational waves. In fact, since the evolution problem for the EFEs is still $1+1$ dimensional, it seems to be only possible hope to achieve exact models of interaction between matter and gravitational radiation. In spite of that, a fair ammount of work must still be done in order to achieve physically reasonable models of collapse to black holes with gravitational wave emission, in cylindrical symmetry. This is due to the pathologies that the simplest models show once a trapped surface forms in the interior. 

It is possible that, for a fair ammount of potentially interesting situations, there will exist no-go results, for example stating that gravitational waves cannot reach infinity due to the presence of a singularity in the exterior. A possibility to circunveil this problem would be to cut the exterior spacetime, removing the singularity, and to match it to another metric. In that case, the gravitational wave spacetime would act as an intermediate region between the collapsing matter and a spacetime background, possibly galactic or cosmological.  


In any case, the cylindrically symmetric solutions can surely continue to give important insights, as an intermediate step, towards the modelling of axially symmetric compact objects in astrophysics.  

\section*{Acknowledgments}

The author thanks R. Vera for his careful reading of the manuscript and for his many suggestions
 and references which greatly improved the paper; J. Senovilla for providing reference \cite{Senovilla};
  the organisers of the VII Black Hole Workshop; FCT project CERN/FP/123609/2011; CMAT, Univ. Minho,
 through FEDER funds COMPETE and FCT funds Est-OE/MAT/UI0013/2014.

\end{document}